\documentclass[10pt,tightenlines,eqsecnum,floats,aps,amssymb,nofootinbib,prd,shownopacs,floatfix,twocolumn,superscriptaddress]{revtex4-2}

\usepackage[a4paper, total={7in, 9in},bottom=20mm,top=25mm]{geometry}
\usepackage{amsmath}
\usepackage{ulem}
\usepackage{tikz}
\usepackage{graphicx}
\usepackage{caption}
\usepackage{subcaption}
\usepackage{dsfont}
\usepackage{braket}
\usepackage[sort&compress]{natbib}
\usepackage{hyperref}
\usepackage{url}    
\makeatletter
\renewcommand{\p@subsection}{}
\renewcommand{\p@subsubsection}{}
\makeatother
\newcommand{\nocontentsline}[3]{}
\let\origcontentsline\addcontentsline
\newcommand\stoptoc{\let\addcontentsline\nocontentsline}
\newcommand\resumetoc{\let\addcontentsline\origcontentsline}

\begin{document}

\title{Effects of quantum geometry on the decoherence induced by black holes}

\author{Max Joseph Fahn}
\email{maxjoseph.fahn@unibo.it}
\affiliation{Dipartimento di Fisica e Astronomia,
Universit\`a di Bologna, Via Irnerio 46,
40126 Bologna, Italy}
\affiliation{INFN Bologna, Via Irnerio 46, 40126 Bologna, Italy}

\author{Alessandro Pesci}
\email{pesci@bo.infn.it}
\affiliation{INFN Bologna, Via Irnerio 46, 40126 Bologna, Italy}

\begin{abstract}
Recently, it has been shown that a quantum system held
in spatial superposition and then eventually recombined
does experience decoherence from black hole horizons, 
at a level increasing linearly with the time 
the superposition has been kept open.
In this, the effects of the horizon have been derived
using a classical spacetime picture for the latter.
In the present note we point out that quantum aspects of the geometry
itself of the quantum black hole could significantly affect
the results.
In a specific effective implementation of the quantum geometry 
in terms
of a minimal length and ensuing minimal area,
it appears in particular that, for selected values of the
quantum of area proposed on various grounds in the literature,
the decoherence induced by the horizon turns out to be
limited to
negligibly small values.
\end{abstract}

\maketitle

The study of decoherence induced by gravity has become a topic of increasing interest in recent years with works focusing for instance on gravitational waves (see e.g. \cite{Blencowe:2012mp,anastopoulos2013master,oniga2016quantum,bassi2017gravitational,lagouvardos2021gravitational,fahn2023gravitationally,Domi:2024ypm,fgSoon,cho2025non}), cosmological setups (see for instance \cite{Boyanovsky:2015tba,boyanovsky2015effective,hollowood2017decoherence,martin2018observational,colas2022benchmarking,brahma2024time}), as well as spacetime curvature and black hole spacetimes (see e.g. \cite{ellis1997quantum,hsu2009black,Danielson:2022tdw,Danielson:2022sga,singh2023decoherence,Gralla:2023oya,Wilson-Gerow:2024ljx,Danielson:2024yru,Li:2025vcm}). The seminal works in \cite{Danielson:2022tdw,Danielson:2022sga,Gralla:2023oya,Wilson-Gerow:2024ljx,Danielson:2024yru} have shown that the mere presence of a horizon
induces decoherence on a quantum system put in a superposition of locations and then recombined,
with total decoherence increasing linearly with the time the system has been held in superposition.

This has also been considered in the perspective of an
extension of a gedankenexperiment from \cite{Baym:2009zu,Mari:2015qva, Belenchia:2018szb} that discusses the compatibility of causality and complementarity in an apparently paradoxical situation in which 
a quantum particle $A$ is held in superposition of locations and eventually recombined
while another particle, $B$, at a distance, probes the (electromagnetic or gravitational)
field sourced by $A$, moving freely under its influence.
If the field sourced by $A$ is quantum, $B$ will entangle with $A$, and a measurement
of the position of $B$ can allow discrimination of the path followed by $A$,
then, from the complementarity principle, implying decoherence of $A$'s state.

While this gedankenexperiment seems to bring in a tension between causality and complementarity
when $A$ recombines and $B$ performs which path in a time so short that they are causally disconnected,
the findings in \cite{Belenchia:2018szb} are that this is in fact not the case: 
when conditions (charge/mass and separation) are such that 
the measurement of the position of $B$ at ideal resolution (dictated by vacuum fluctuations)
allows for a discrimination of the path of $A$, the recombination of the latter in a so short time 
will be accompanied by emission of a quantum of (electromagnetic or gravitational) radiation, 
that destroys the coherence of $A$'s state,
without any need of getting information from $B$.

In \cite{Danielson:2022tdw,Danielson:2022sga} this gedankenexperiment is extended by placing $B$ behind a horizon.
In this situation, 
$A$ can be recombined arbitrarily slow, 
since it no longer needs to recombine quickly in order not to get information from $B$.
This clearly challenges in principle the resolution of the paradox along the lines above.
There is a resolution, however, proposed in \cite{Danielson:2022tdw,Danielson:2022sga}:
The mere presence of a horizon induces decoherence on $A$.
This effect is caused by the soft modes (i.e. the modes of the quantized field under consideration with very low frequencies) that pierce the horizon. This leads to a decoherence in $A$'s superposition that grows linearly in the time $\mathcal{T}$ the superposition is kept open for large enough times. For the electromagnetic field, the case on which we focus in this article, the mean number $\langle N \rangle$ of radiated particles through the horizon, responsible for the horizon-induced decoherence, was shown in \cite{Danielson:2022tdw,Danielson:2022sga} to be proportional to 
\begin{equation}\label{eq: Wald proportionality}
    \langle N \rangle \propto \frac{M^3 q^2 d^2}{D^6} \mathcal{T}\,,
\end{equation}
where $M$ is the mass of the black hole, $q$ the electric charge of the particle, $d$ the spatial separation of A's superposition and $D$ its distance from the horizon. Furthermore, \cite{Danielson:2022sga} considers more general spacetimes with Killing horizons, such as Rindler or cosmological, finding a similar effect.

These findings were confirmed in \cite{Gralla:2023oya}, extended to discuss an observer on the symmetry axis of a Kerr black hole and the proportionality factor lacking in equation \eqref{eq: Wald proportionality} was determined. While these works have followed the ``global approach" by considering the radiation in the entire system, the same results are obtained in \cite{Wilson-Gerow:2024ljx,Danielson:2024yru} by only evaluating quantities available at $A$'s position (``local approach"). The same behavior is found in \cite{Kawamoto:2025kfu} in a holographic context,
and in \cite{Biggs:2024dgp} by treating the black hole as a ``black'' quantum system at finite (Hawking's) temperature interacting through its multipole moments, followed in the same spirit by \cite{Li:2025vcm} where near-extremal Reissner-Nordström black holes were discussed and a partial enhancement of the decoherence was found for specific configurations of the black hole. 
All these works have in common that they consider a classical, or effective classical-limit, spacetime geometry on which the quantum objects, i.e., the particle $A$ and the quantum field it interacts with, are placed, and correspondingly also work with an effective classical geometry for the horizon.

Given the prospect of an actual quantization of spacetime existing in several candidate and effective theories for quantum gravity, e.g. \cite{Bekenstein,Ashtekar,Hod,Maggiore,Barbero2,KriPer,DICE}, in this letter we pose the question of the influence of effects of a quantum geometry on the decoherence process described above. 
The quantization of geometry we are thinking at here
is meant as implemented in an effective way 
by use of the existence of a minimal area 
variation of black hole horizons. 
This goes back to the original assumption of Bekenstein \cite{Bekenstein} 
and has been reproposed on different grounds and with different explicit values for it since then, 
and has been also shown to be derivable from the existence of a minimal length in \cite{KotE, KotF, KotI,KriPer,DICE}.

It will turn out that this feature of a quantized geometry will in particular prevent modes whose energies lie below that corresponding to the minimal area change to be absorbed by the black hole and hence to enter it, similarly as transitions inside atoms can only happen if the photons have a suitable (in the analogy used here a sufficiently large) frequency.
As we will see, the consideration of simply just one single 
minimum-area step can give already, for the values of the quanta
of area proposed so far, the bulk of the effect. 
Given that the decoherence process described in \cite{Danielson:2022tdw,Danielson:2022sga} is caused by very low frequency radiation, we obtain possibly quite a significant change in the behavior of the decoherence caused by the presence of a horizon with quantum geometry.
In what follows, we first briefly introduce the discussions for horizons with classical geometries and then show how they can be modified to take into account a quantization of the horizon geometry by the introduction of a minimal area. 

Let us first briefly review some
results for the case of a classical stationary horizon in order to set the
stage for horizons with quantum geometries.
We will focus here on the local approach,
specifically on the derivation in \cite{Wilson-Gerow:2024ljx},
and discuss the electromagnetic case. 
We consider a charged particle $A$, in a stationary lab in
a black hole spacetime which, to illustrate the case,
we take to be Schwarzschild.
The particle
is put in a quantum superposition of two locations,
described by the orthogonal states $\ket{\psi_L}$ and $\ket{\psi_R}$,
held in this superposition for a time $\mathcal{T}$,
and then eventually recombined. 
We assume that the associated electromagnetic currents 
$j^\mu$, $j^\mu_L$ and $j^\mu_R$,
with $j^\mu_L = j^\mu_R = j^\mu$ at early and late times,
are essentially
classical,
i.e., quantum fluctuations are negligible compared to their expectation values and do not exhibit
appreciable radiation reaction, at all times.
The sourced quantum field is then in the coherent states
$|A^\mu\rangle$, $|A^\mu_{L/R}\rangle$ \cite{Glauber}.
While delocalized, the state evolves as
\begin{eqnarray}
    |\chi\rangle = \frac{1}{\sqrt{2}} 
    \big(|\psi_L\rangle \, |A_L\rangle
    + |\psi_R\rangle \, |A_R\rangle\big)
    \ ,
\end{eqnarray}
where we dropped spacetime indices in the field states for ease of notation.
This can be usefully expressed in terms of the density matrix $\rho_{\psi A}   = \ket{\chi}\bra{\chi}$.
The reduced density matrix for the particle alone is obtained by tracing out the field
and reads
\begin{eqnarray}\label{rho}
\rho_\psi
&=&
\frac{1}{2} \Big( |\psi_L\rangle\langle\psi_L| +\langle A_R|A_L\rangle |\psi_L\rangle\langle\psi_R|
\nonumber
\\
&&
\phantom{\frac{1}{2}}
+ 
\langle A_L|A_R\rangle |\psi_R\rangle\langle\psi_L|
+ |\psi_R\rangle\langle\psi_R|\Big)
\nonumber
\\
&=&
    \begin{pmatrix}
        \rho_{LL} & \mathcal{F}[L,R]\,\rho_{LR}
        \\
        \mathcal{F}[R,L]\,\rho_{RL} & \rho_{RR}
    \end{pmatrix}
    \ ,
\end{eqnarray}
where $\rho_{LL}$ and $\rho_{RR}$ are the probabilities that the particle is found in one of the two branches.
The off-diagonal elements denote interference probabilities and $\mathcal{F}$ is the Feynman-Vernon influence
functional \cite{Feynman:1963fq}.
Its absolute value is defined as 
\begin{equation}\label{deco_2}
 |\mathcal{F}[L,R]|
 =
 |\mathcal{F}[R,L]|
 \equiv 
 |\mathcal{F}| = e^{-\frac{1}{2}\mathds{D}}
 \ ,
\end{equation}
with the decoherence functional $\mathds{D}\geq 0$.
It can be readily seen that the latter, if non-zero, damps the interference probabilities and thus leads
to decoherence of A's system.
We assume here that the experimenter has been
able to perform the separation coherently,
so that $\mathds{D} = 0$ for the superposition just opened. 
The decoherence functional encodes the influence of the environment, in the present case
of the quantum field (which sees the horizon and thus contains the information of its presence).
It turns out to be the number $\langle N \rangle$ of 
(entangling) particles sourced by $j_L-j_R$, i.e.,
by a matching electric dipole replacing the superposition. 

The idea in~\cite{Wilson-Gerow:2024ljx} is to study the state of the particle 
using the theory of open quantum systems
(see e.g.~\cite{breuer2002theory} for an introduction)
and model the particle's spatial superposition in terms of the Rindler frame of an accelerating electric dipole.
Our focus in this work is to apply it
near the horizon of the black hole -- because this is where we would expect the strongest decoherence effect --
as the local Rindler horizon approximating 
the actual stationary horizon.

The dipole
is created during some  proper time ${\cal T}_1$, then kept open for an interval ${\cal T} \gg {\cal T}_1$
and closed afterwards during a time interval which we also choose to be ${\cal T}_1$.
The expression for the decoherence functional $\mathds{D}$ can be obtained from the Feynman-Vernon influence functional, under the assumptions that  higher than second order correlation functions of the electric field vanish and that the influence functional can be evaluated along the classical trajectory of $A$'s particle, see \cite{Wilson-Gerow:2024ljx} for details, to be\footnote{We work in units with $\hbar=c=G=1$ and restore the constants when discussing numerical values.}
\begin{align}\label{eq:defdecfctft}
    \mathds{D} &= \frac{q^2}{2} \int_{-\infty}^\infty d\tau\, d\tau'\; \epsilon^A(\tau)\; \epsilon^B(\tau') \langle\{E_A(X(\tau)),E_B(X(\tau'))\}\rangle \nonumber\\ &=\frac{q^2}{2} \int_{-\infty}^\infty \frac{d\Omega}{2\pi}\; S_{AB}(\Omega) \; \tilde{\epsilon}^A(-\Omega) \, \tilde{\epsilon}^B(\Omega)\,,
\end{align}
where $\epsilon^A(\tau)$, $A=1,2,3$, denote the components of the separation vector in the orthonormal lab frame evaluated at proper time $\tau$, and $E_A(X(\tau))$ the electric field in the lab frame evaluated on the Rindler trajectory with proper acceleration $a$,
\begin{equation}
    X^\mu(\tau) = \frac{1}{a} (\sinh(a\tau),0,0,\cosh(a\tau))
\end{equation}
(having taken the $X^3$-axis
along the acceleration direction).

In the second line of equation \eqref{eq:defdecfctft}, a tilde denotes Fourier transform with respect to proper time and $S_{AB}(\Omega)$ the Fourier transform of the two-point correlation function of the electric field which turns out to be
\begin{equation}
    S_{AB}(\Omega) = \delta_{AB} \frac{\Omega^3 + a^2\Omega}{3\pi\epsilon_0} \coth\left( \frac{\pi \Omega}{a} \right)\,.
\end{equation}
The term in the numerator that goes linear in $\Omega$ is not present for a particle at rest in a thermal environment at Hawking temperature, as was discussed in \cite{Wilson-Gerow:2024ljx}. As shown there, it is the crucial term that leads to the decoherence caused by the horizon: Assuming the separation of $A$'s superposition, $\epsilon^A(\tau)$, to be constant, $\epsilon^A$, for a long proper time $\mathcal{T}$ and zero otherwise, results in its Fourier transform being strongly peaked around $\Omega=0$ and hence the integration in frequency domain in equation \eqref{eq:defdecfctft} can be approximated by only considering the modes around $\Omega=0$ (see e.g. \cite{Wilson-Gerow:2024ljx}):
\begin{align}\label{eq:Dfctapro1}
    \mathds{D} 
    &\approx \frac{q^2}{2} S_{AB}(0) \int_{\mathds{R}} \frac{d\Omega}{2\pi} \; \tilde{\epsilon}^A(-\Omega) \tilde{\epsilon}^B(\Omega)\nonumber\\ &= \frac{q^2}{2} S_{AB}(0) \int_{\mathds{R}} d\tau \; {\epsilon}^A(\tau) {\epsilon}^B(\tau) \nonumber\\ &\approx \frac{q^2}{2} \epsilon^A \epsilon^B \, {\cal T} \, S_{AB}(0) = \frac{q^2d^2 a^3}{6 \pi^2
    \epsilon_0} \, {\cal T} \,.
\end{align}
In accordance with the results from \cite{Danielson:2022tdw,Danielson:2022sga,Gralla:2023oya,Danielson:2024yru,Biggs:2024dgp}, this leads to decoherence induced by the horizon,
with a linearly increasing decoherence functional for sufficiently large
times $\mathcal{T}$.
The main contribution of the decoherence effect comes from the two-point correlation function at $\Omega\approx 0$, that is from the very low frequency modes of the electric field. In this discussion, in particular the decoherence caused by the opening and closing of the superposition during $\mathcal{T}_1$ and from radiation going to infinity is neglected, as it can be bounded from above while the decoherence caused by the horizon can become arbitrarily large (i.e. arbitrarily close to unity) when waiting long enough in $\mathcal{T}$.

A crucial ingredient in this calculation, which is present in all the works \cite{Danielson:2022tdw,Danielson:2022sga,Gralla:2023oya,Wilson-Gerow:2024ljx,Danielson:2024yru,Biggs:2024dgp,Li:2025vcm} in one or another way, is the assumption that the entire spectrum of electric field modes can pierce the horizon (in the global point of view) and that the presence of the horizon modifies the spectrum of the electric field at all frequencies (in the local point of view). 
If at the very small scales we are in presence of a quantum geometry, this can be challenged.

Taking indeed as basic element of quantumness of the geometry
the existence of a fundamental limit length 
(as generically expected from joining general relativity with basic quantum theory) one gets a minimal step $A_0$ for
area transverse to horizon's generators \cite{Padmanabhan:2015pza, QuantumMetricNull, KriPer},
and 
the changes in the horizon's total area, if any,  are at least by this value.
Due to energy conservation, this is only possible if the horizon absorbs enough energy $E$ for the transition from the mass state corresponding to the area $A$ before the absorption to the new mass state corresponding to $A+A_0$.

As a consequence, the minimal step in area implies a minimal energy and frequency a particle needs to possess in order to be able to interact with the horizon, in other words to see the horizon. For a horizon corresponding to a Schwarzschild black hole of initial mass $M$, the minimal frequency (in asymptotic time) is

\begin{equation}
    \omega_0 = \frac{\kappa \, A_0}{8 \pi}\,,
\end{equation}
with surface gravity $\kappa=\frac{1}{4M}$, 
with $A_0 = 8 \pi$ 
for the original quantum-of-area proposal \cite{Bekenstein} and for \cite{Maggiore}, and $A_0 = 4 \ln 3$
for \cite{Hod, Barbero2} (other proposed selected values
are approximately in this range).

Clearly the prediction of such kind 
of effects involves a big assumption since it refers to scales
extremely small, for which we have no guarantee that physics
works the same as at (much) larger scales.
On the other hand the effects we are considering appear to depend
critically on what happens at these tiny scales, this is our point, and an assumption is needed precisely because
there is no direct experimental control so far.

The existence of a limit length is quite a generic prediction at least at an effective level, from very
diverse quantum approaches to gravity \cite{Garay:1994en}. 
It regulates divergences in quantum field theory and
singularities in general relativity \cite{Deser:1957zz}, and DeWitt has shown
that an effective action for gravity devoid of singularities
leads to a propagator in which to the quadratic distance is added a constant which acts as a residual distance when
two points go to coincide \cite{dewitt1964gravity, dewitt1981approximate}, sort of zero-point-length (see
Padmanabhan on this \cite{padmanabhan1985physical}).
This zero-point length can be implemented in the metric description in a Lorentz-invariant manner \cite{KotE, KotF, KotI}, 
and this is the prescription we use in this study,
also in view of the fact that models involving modification
or violation of Lorentz invariance appear to be severely
constrained by available data or suffer from inconsistencies and
it seems better to give up with locality rather than
Lorentz invariance \cite{garay1998spacetime, Doplicher:1994tu, Sorkin:2007qi, giddings2006locality}.
In spite of these virtues we have to keep in mind that
it is an assumption, and that the results we now go to present 
do hinge on it.

As mentioned, 
we use the Rindler horizon above
as the near-horizon limit of the Schwarzschild black hole.
This comes with the correspondence
$\{\Omega,a\}\longleftrightarrow\{\omega,\kappa\}$
with $\Omega = \frac{a}{\kappa} \, \omega$
in the integral of equation \eqref{eq:defdecfctft} above,
when passing from asymptotic to Rindler time
in the near-horizon limit
(cf. \cite{Gralla:2023oya}). 

The quantization of the area geometry can be implemented in the calculation of the decoherence functional equation \eqref{eq:defdecfctft} by excluding modes with frequency below the threshold frequency $\Omega_0 < \frac{a}{\kappa}\omega_0$ \cite{KriPer,DICE} (cf. also \cite{Emparan:2025qqf}).

As a consequence, the approximation used in equation \eqref{eq:Dfctapro1} in order to evaluate the decoherence functional needs to be reconsidered, as the very low frequency modes with $\Omega\approx 0$ cannot (or, depending on the size of $\Omega_0$ only partially) contribute to the decoherence functional any more.

To discuss the effect of this modification, let us first give some fairly general expressions for the quantities in equation 
\eqref{eq:defdecfctft}
for 
small $\Omega_0 \lesssim a$.
In this case, we can obtain for the separation vector $\epsilon^A(\tau)$,
\begin{align}\label{eq:dipolmodel}
    \epsilon_1(\tau) = \begin{cases}
d & \text{for } -\frac{\cal T}{2} \leq \tau \leq \frac{\cal T}{2}\\
0 & \text{else}\,, 
\end{cases}
\end{align}
where we chose $\epsilon_2(\tau)=\epsilon_3(\tau)=0$. This implies for equation \eqref{eq:Dfctapro1}:
\begin{align}\label{eq:Decfct}
    \mathds{D} \approx \frac{2q^2 d^2}{3\pi^2\epsilon_0}\int_{\Omega_0}^{\overline{\Omega}} d\Omega\;  \frac{\Omega^2 + a^2}{\Omega} \sin^2\left(\frac{\Omega \mathcal{T}}{2} \right) \coth\left( \frac{\pi\Omega}{a} \right)\,,
\end{align}
where the case of a classical horizon in equation \eqref{eq:Dfctapro1} is obtained for $\Omega_0=0$.
We have also included
an upper frequency limit $\overline\Omega> a$ to
express the fact that the relevant range of integration can be effectively regarded as finite by a suitable choice of the transients.

Let us now investigate the effect of a non-zero lower frequency cut $\Omega_0$, arising from a horizon with quantized geometry, on the decoherence functional. From the form of the integrand in equation \eqref{eq:Decfct} it is evident that the dependence on $\mathcal{T}$ becomes constant for values of $\Omega$ that fulfill $\Omega\mathcal{T}\gg 1$, as in that case the $\sin^2$ term will contribute fast oscillations that can be averaged with a factor of $\frac{1}{2}$, so values of $\Omega \gg \frac{1}{\mathcal{T}}$ only contribute a constant to the decoherence. This however implies that the dependence in $\mathcal{T}$ arises from small values of $\Omega$.

\begin{figure}[!hbtp]
  \centering
  \includegraphics[width=0.43\textwidth]{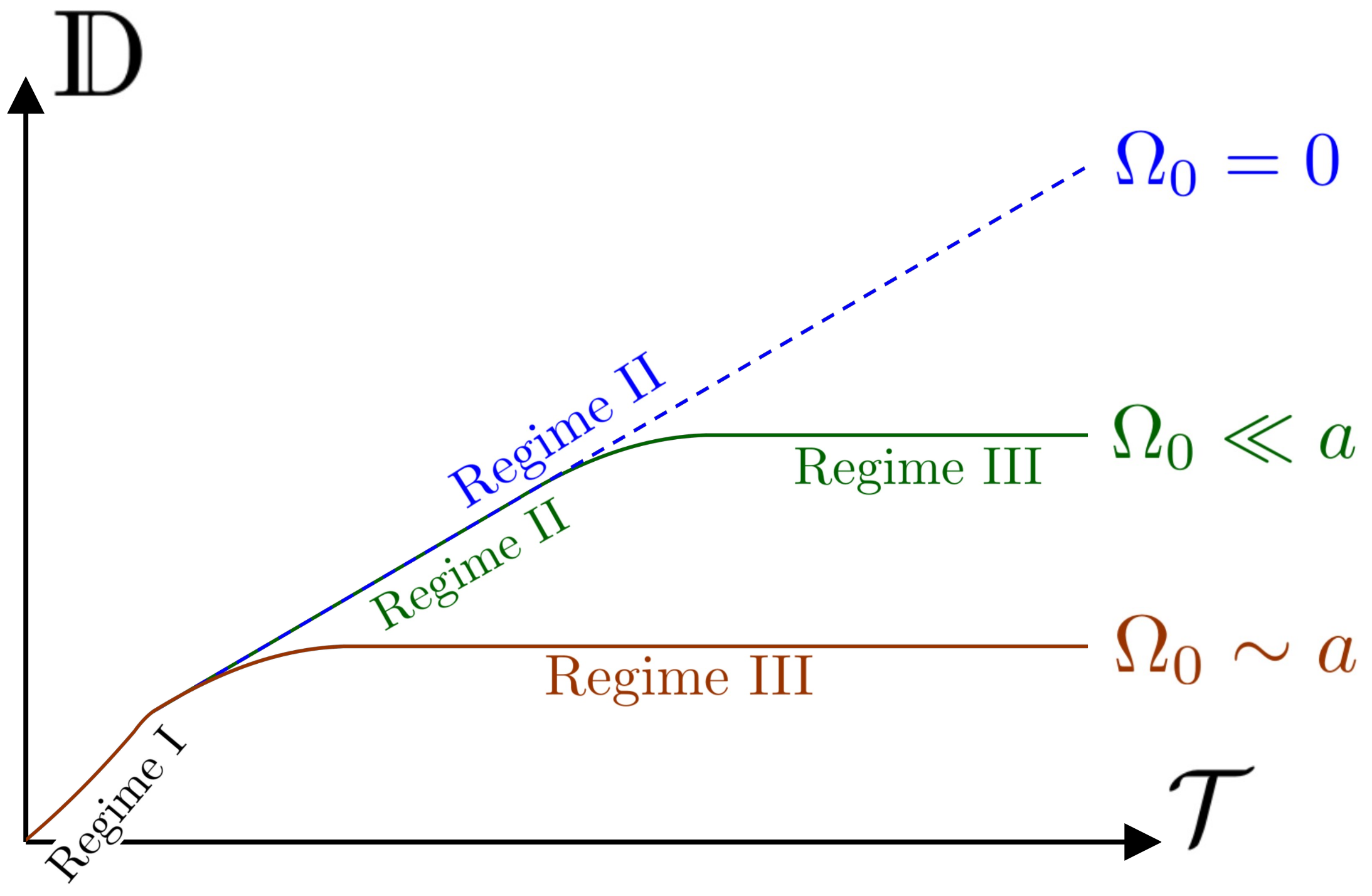}
  \caption{Qualitative sketch of the behavior of the decoherence functional $\mathds{D}$ for different sizes of cutoffs $\Omega_0$ or no cutoff at all ($\Omega_0=0$). The saturation values obtained in Regime III for the cutoffs proposed in the literature that are of the order of $\Omega_0\sim a$ are many orders of magnitude lower than unity, see equation \eqref{eq:dsatval}, thus rendering the decoherence negligibly small. The behavior at the transitions between the different regimes is not accurately resolved in this sketch as the main purpose is a visualization of the three different regimes.
  }\label{fig:plot}
\end{figure}

We are now interested in determining the dependency of $\mathds{D}$ on $\mathcal{T}$ in equation \eqref{eq:Decfct}. The qualitative behavior discussed in what follows is visualized in figure \ref{fig:plot}.
We shall identify the following regimes:\\
- \underline{Regime I}: For very small $\mathcal{T}$, i.e. such that $\overline\Omega \mathcal{T} \ll 1$, one can approximate $\sin^2\left( \frac{\Omega\mathcal{T}}{2}\right) \approx \frac{\Omega^2}{4} \mathcal{T}^2$ and obtains thus a dependency of $\mathds{D} \propto \mathcal{T}^2$. This case is however of limited interest as the main focus in \cite{Danielson:2022tdw,Danielson:2022sga,Gralla:2023oya,Wilson-Gerow:2024ljx,Danielson:2024yru} is on the case $\mathcal{T}\gg \mathcal{T}_1 \sim 1/\overline{\Omega}$.
\\
- \underline{Regime II}: For larger $\mathcal{T}$, at a certain point we will reach a regime where $\mathcal{T} \gg \frac{1}{a}$, that is the dependency on $\mathcal{T}$ in $\mathds{D}$ only comes from frequencies $\Omega\ll a$. In this regime we can approximate $\coth\left(\frac{\pi \Omega}{a}\right)\approx \frac{a}{\pi \Omega}$ and obtain, if the cut is so small such that in that regime $\Omega_0 \mathcal{T}\ll 1$ still holds:
    \begin{align}\label{eq:linreg}
        \mathds{D} &\approx \text{const} + \frac{2q^2d^2}{3\pi^2 \epsilon_0} \int_{\Omega_0}^{\widehat{\Omega}}d\Omega\; \frac{a^3}{\pi \Omega^2} \sin^2\left( \frac{\Omega\mathcal{T}}{2}\right)  \nonumber\\ &\approx\text{const} + \frac{q^2 d^2 a^3}{6\pi^2 \epsilon_0}\mathcal{T}\,,
    \end{align}
    where $\frac{1}{\mathcal{T}} \ll \widehat{\Omega} \ll a$.
    This is the case which for $\mathcal{T}$ large enough gives the same linear behavior as horizons with classical geometry, as discussed above in equation \eqref{eq:Dfctapro1}, with a same slope. 
    When $\Omega_0=0$, this is the final stage of discussion.
    If $\Omega_0>0$ however, the behavior of the decoherence functional can change again, see the next bullet point.
    Note that when the cut is too large, i.e. when $\Omega_0$ is of the order of $a$,
    as it is indeed the case for some of the values proposed in the literature mentioned above, 
    then this regime can be of very limited extent 
    (if any).\\
- \underline{Regime III}: For $\mathcal{T}$ large enough to have $\Omega_0 \mathcal{T} \gg 1$, the contribution from $\sin^2$ can be approximated under the integral by a factor of $\frac{1}{2}$ due to the fast oscillations. In this regime the decoherence runs into a value independent of $\mathcal{T}$ which we call saturation value $\mathds{D}_{\text{sat}}$.

Without the presence of a cut, the last regime is never reached as arbitrarily low modes are included in the integration that always fulfill $\Omega \mathcal{T} \ll 1$. With the inclusion of a cut these modes are excluded, such that a saturation is reached. As is visible from this qualitative discussion, the larger the cut, the faster in $\mathcal{T}$ the saturation regime, characterized by $\mathcal{T}\gg \frac{1}{\Omega_0}$, will be reached.

These considerations show that the decoherence induced by horizons with classical geometries can be strongly modified when considering horizons possessing quantum geometries in the way implemented above. Given this result, two main questions remain to be answered. First, the order of magnitude of the saturation value depending on the cut: If it is close to unity for typical cut values suggested in the literature, then the modification caused by the inclusion of quantum horizon geometries instead of classical ones would not really affect the results.
Second, if the saturation value is sufficiently low, the original paradox in the gedankenexperiment seems to be posed again: Then, $B$ behind the horizon seems to be able to discriminate the path in $A$'s superposition but $A$ might not decohere any more. We provide an answer to the first question in what follows and respond to the second at the end.

We have seen that for cuts small enough, the decoherence functional shows a linear increase for times $\mathcal{T}\ll \frac{1}{\Omega_0}$. And runs into a saturation value for $\mathcal{T}\gg \frac{1}{\Omega_0}$.
From a very rough estimate, the order of magnitude of this saturation value is hence provided from equation \eqref{eq:linreg} 
by 
\begin{equation}\label{eq:estDsat}
    \mathds{D}_{\text{sat}} \approx \frac{q^2 d^2 a^3}{6\pi^2 \epsilon_0} \frac{1}{\Omega_0}\,.
\end{equation}
In our accompanying paper \cite{FahnPesci}, an analytical derivation of the exact form of the saturation value is presented, the value being slightly lower but of the same order of magnitude as the one obtained by the rough estimate here. If the cut is of the order of $a$, then the estimate in equation \eqref{eq:estDsat} at least provides an upper bound for the real value of $\mathds{D}_\text{sat}$.

Let us extract a value of $\mathds{D}_\text{sat}$ for possible settings for the gedankenexperiment.
In the near-horizon limit,
the acceleration $a$ is related to the proper distance $D$ of the particle from the horizon by $a=\frac{1}{D} + \mathcal{O}(D)$ 
\cite{Gralla:2023oya}. 
This yields, with all constants reinserted and assuming $q=e$,
\begin{align}\label{eq:dsatval}
    \mathds{D}_\text{sat} &\lesssim \frac{q^2}{6\pi^2 \hbar c \epsilon_0} \left(\frac{d}{D}\right)^2 \frac{1}{c\Omega_0/a} \nonumber\\ &\approx 1.5\cdot 10^{-3} \left(\frac{d}{D}\right)^2 \frac{1}{A_0/(8 \pi l_p^2)}\,.
\end{align}

This shows that, even going very near to the horizon,
values $\Omega_0/a = A_0/(8 \pi) \ll 1$
would be required in order to produce a saturation value not negligibly small.
Hence the linear behavior of $\mathds{D}$, that remains forever in $\mathcal{T}$ for horizons with classical geometries,
would be replaced by a saturation
value which appears quite negligible
for the minimal area values proposed in the literature.
Depending on the specific detailed properties of the creation and recombination of the superposition, this value can even be comparable to the decoherence caused by the modes going to infinity, that were discussed e.g. in \cite{Belenchia:2018szb,Danielson:2022tdw,Danielson:2022sga}.
At this point, one might worry about the size of the constant in equation \eqref{eq:linreg} adding to the saturation value of the decoherence in equation \eqref{eq:estDsat} arising from the $\Omega$-integration for values $\Omega > \widehat{\Omega}$. 
By suitable choice of the transients,
one expects this to be $\ll 1$ and this is indeed 
what is used in the works
\cite{Danielson:2022sga, Danielson:2022tdw, Gralla:2023oya, Wilson-Gerow:2024ljx, Danielson:2024yru} and appears also enough for our needs here
(a different question, which is how this value compares 
with the saturation value 
of equation \eqref{eq:estDsat},
is considered in \cite{FahnPesci}).

Hence to summarize, while the mere presence of a horizon with classical geometry induces decoherence on a particle in a spatial superposition, 
horizons with quantum geometries implemented by the
existence of a minimal area do not perceptibly decohere the spatial superpositions unless the minimal area is many orders of magnitude smaller than the Planck length squared. 

This now opens once again the initial question on how causality and complementarity can both coexist when putting $B$ behind a horizon with quantized geometry. 
Indeed, we have now that $A$'s superposition remains
almost coherent due to the negligible amount of decoherence
induced by the horizon, in contrast with the proposed resolution
mentioned at start.
Is this now finally really a way to contradict the coexistence
of complementarity and causality?
Apparently it is not.
Since the low frequencies that would cause $A$'s superposition
to decohere can not enter the black hole anymore,
no relevant information on the superposition 
can indeed reach $B$, thus making it impossible for $B$
to entangle in a relevant way with $A$, thus inhibiting
which path discrimination.

We wish to thank R. Casadio for comments and suggestions.
This research was partially supported by INFN grant FLAG.
We would like to acknowledge the contribution from
the COST action CA23130, and MJF also from the COST action
CA23115.

\bibliography{references.bib}

\end{document}